\documentclass[11pt]{article}

\usepackage{fullpage}

\usepackage{url}

  \usepackage[pdftex]{graphicx}
    \graphicspath{{./figures/}{./plots/}}
  \DeclareGraphicsExtensions{.pdf,.jpeg,.png}
\usepackage[cmex10]{amsmath}

\usepackage{algpseudocode}

\usepackage{amsmath}
\usepackage{relsize}

\usepackage{longtable}

\usepackage{amsmath,amssymb}

\newcommand{\dc}{\text{{\bf dec}}}
\newcommand{\ec}{\text{{\bf ec}}}
\newcommand{\slpec}{\text{{\bf slp-ec}}}

\newcommand{\slpdeg}{\text{{\bf slp-deg}}}
\newcommand{\ddeg}{\text{{\bf dd}}}

\newcommand{\slpfreq}{\text{{\bf slp-freq}}}
\newcommand{\dfreq}{\text{{\bf df}}}

\newcommand{\tf}{\text{tf}}

\title{Detecting and Summarizing Emergent Events in Microblogs and Social Media Streams by Dynamic Centralities}

\author{Neela Avudaiappan, Alexander Herzog, Sneha Kadam, Yuheng Du, Jason Thatcher, Ilya Safro\\ School of Computing, Clemson University \\ 
    Clemson, South Carolina 29634\\
    Emails: (navudai, aherzog, skadam, yuhengd, jthatch, isafro)@clemson.edu}

\begin{document}
%

\maketitle

\begin{abstract}
%
Methods for detecting and summarizing emergent keywords have been extensively studied since social media and microblogging activities have started to play an important role in data analysis and decision making. We present a system for monitoring emergent keywords and summarizing a document stream based on the dynamic semantic graphs of streaming documents. We introduce the notion of dynamic eigenvector centrality for ranking emergent keywords, and present an algorithm for summarizing emergent events that is based on the minimum weight set cover. We demonstrate our system with an analysis of streaming Twitter data related to public security events.

\end{abstract}

\section{Introduction}

Today, federal and state security organizations invest enormous efforts in detecting and monitoring public security events such as social protests, migrant crisis, and terrorist attacks. Unfortunately, emergent security risks associated with such events are not timely detected and monitored. Among the most enlightening examples are the New Year's Eve coordinated sexual assaults in Germany \cite{roberts2016europe,eddy2016reports}, Boston marathon bombing \cite{von2016mental,wormwood2016threat}, and massive terrorist attacks in France \cite{franceattacks}. As the details of such events begin to emerge, authorities, and citizens around the world are storming social media. This includes both the condemnations and defense sides. After the self-proclaimed Islamic State (ISIS) claims responsibility for the violence, the social media is abuzz with both the condemnations of and solidarity with the extremist group and its actions. 

Another major problem with the blogosphere and social media highlighted in many recent studies discusses the enormous volume of online propaganda, suggesting that acts of terror may help the terrorist group attract supporters and conduct its recruitment. All these and many other reasons put detecting and monitoring of emergent public safety events in streaming data among the top priorities of federal and state authorities. It has been shown multiple times that relevant information can appear in Twitter much faster than the authorities begin to respond. For example, in the 2015-2016 New Year's Eve coordinated sexual assaults in multiple cities in Germany, people started to report via Twitter much earlier than the local authorities started to react. It has been discussed in media that the reporting of such incidents in one city would have been helpful in preventing similar events other cities. Trend detection is also an important task for other types of streaming data monitoring, such as search engine query monitoring \cite{Golbandi:2013:EST:2433396.2433435}, climate related spatial temporal data analysis \cite{Sonali2013212}, and social media event analysis \cite{benhardus2013streaming}\cite{saha2012learning}. 

The methods for implementing trend detection are often task oriented and data dependent. 
In search engine query monitoring, trend detection techniques identify the rising queries that reflect the user's attention at the moment. In \cite{Golbandi:2013:EST:2433396.2433435}, Goldbani et al. advanced the basic search trend detection algorithm proposed by Dong et al. in \cite{dong2010towards} and reduced the latency of the detection algorithm by 20 minutes. They used linear regression to predict the future query counts. By feeding these predictions into Dong et al.'s model, the detection procedure was significantly accelerated. The query data usually consists of short text phrases and query counts of each search term are often the most crucial feature for search trend detection. Both methods are based on query counts, where correlations between search terms are neglected. These methods work great for news-like, fast emerging search data.

Often, trend detection techniques for social media use approaches from traditional statistical data trend detection such as those in climate computations. In \cite{Sonali2013212}, Sonali and Kumar applied both traditional statistical tests and cutting-edge trend analysis method \cite{csen2011innovative} to study the trend of  temperature changes in India. The detection results showed evident increasing trend in temperature over the past three decades in India. While statistical methods such as Mann-Kendall (MK) \cite{kendall1948rank} test and Sen's cartesian plane trend analysis \cite{csen2011innovative} are used to detect this trend, certain preconditions exist. For example, in the MK test, the data points are assumed to be independent; and in Sen's test different time scales are needed (yearly, monthly, seasonal) to draw the correct conclusion. Climate data, such as temperature and precipitation, are not as vast in speed and volume as social data or web search data. Hence, the detection latency is less of a concern. Moreover, the trend detection on climate data is focused largely on trends over different time spans and are rarely on abrupt changes.

In social media, trend detection is critical for identifying the changing and emerging themes from the vast amount of text streams \cite{Mathioudakis:2010:TTD:1807167.1807306} \cite{benhardus2013streaming} \cite{saha2012learning}. Due to the fast changing nature of social data, trend detection methods are often built around abrupt changes from real-time data streams. Mathioudakis and Koudas \cite{Mathioudakis:2010:TTD:1807167.1807306} built a Twitter Monitor tool which detects erupting keywords from live Twitter stream and group them into small groups based on the co-occurrences of keywords in recent tweets. A trend is then represented by a group of keywords that can be manually tagged by users for monitoring purposes. Benhardus and Kalita's work \cite{benhardus2013streaming} on Twitter data determines if a keyword should be selected into a trending topic by checking its term frequency-inverse document frequency (TF-IDF) value. The eventual trending topic being identified contains all the words that have a TF-IDF value greater than a certain threshold. Since there is no topic-word probabilistic model involved in Benhardus and Kalita's work, their trend detection approach is mainly based on word frequency and associated thresholds. Saha and Sindhwani \cite{saha2012learning} introduce topic-word and topic-document association model and build a nonnegative matrix factorization (NMF) framework that separates evolving topics and emerging topics from Twitter data in real time. At the end of each time window, top $K$ most emerging topics are generated as the trending topics. The rest of the topics are generated as the smoothly-changing evolving topics. No explicit threshold is needed for this approach and it outperforms the threshold based method used in Allan and James' \cite{allan2012topic} work in both the computational time and precision. Despite the differences in statistical methods used in \cite{saha2012learning} and \cite{benhardus2013streaming}, both studies relied their trending keyword recovery process on the frequency of keywords in a given time range. 

Several methods have been developed in the past to extract keywords and summarize text using  semantic graphs in which keywords are represented by nodes, and edges represent one of the measures based on the co-occurrence of keywords in the same documents. Graph based approaches have been used for extracting important keywords using centrality measures such as closeness centrality \cite{matsuo2001keyworld} that measures how well a keyword is connected to all other keywords in the graph by taking into account the shortest path distance between them. 

Another cohort of relevant centrality measures are based on the flows between nodes \cite{brandes:flow} as it models the amount of possible information content that can be transferred between nodes. In some methods, the centrality index is replaced with the corresponding vitality index that measures the effective changes in the network property after removal of a node or an edge \cite{book:brandes}. In \cite{erkan2004lexrank}, a problem of text summarization is solved by finding the most important sentences, using eigenvector centralities. A connectivity matrix constructed using intra-sentence cosine similarity which is then used as the adjacency matrix of the graphical representation of sentences. In \cite{erkan2004lexrank},  the centrality measure is used in context of sentences as opposed to words in the sentences in other works. Some methods have used a hypergraph structure to maintain information regarding words \cite{bellaachia2014hg}. In this structure hyperedges are documents which may contain many vertices. The vertices correspond to words. The hyperedges maintain temporal information about words which is used in centrality measurement for dynamically changing graphs. Some methods use polynomial regression to predict centrality values in future \cite{kim2012centrality}. 



\noindent {\bf Our Contribution} 

Social media is a rich corpus of information that usually contains crucial data on real time trending events. Twitter is one of the most popular social media websites that has witnessed a huge hike in its usage in the recent years. The diversity and volume of users provide different perspectives through tweets on news events around the world. Analyzing tweets can hence unfold useful information on impactful events. The analysis is challenging because of large volume of data and noise. This introduces us to the problem of finding the most relevant tweets leaving behind the less important ones. Also, the solution has to be scalable and capable of processing real time streaming data.

We introduce a simple novel method to detect emergent keywords in data streams that is based on the analysis of \emph{dynamic} semantic graphs. In contrast to many other semantic \emph{static} graph-based approaches, we introduce a notion of dynamic node centrality to measure the emergent importance of keywords. We generalize the well known frequency, degree, and eigenvector centralities into corresponding dynamic versions, and demonstrate their performance on a stream of Twitter data related to two public security events, namely, Boston marathon bombing, and protests in Baltimore. Furthermore, we introduce an algorithm for the data stream summarization and demonstrate it on the same data sets. The summarization approach is based on the minimum weighted set cover algorithm applied on the semantic graph of the dynamically highly ranked keywords. We compare the quality of all methods. The implementation is freely available at \cite{dyncentr-impl}.


\section{Modeling approach}
In the heart of the proposed modeling approach lies a dynamic semantic graph of keywords in which the nodes and undirected edges correspond to keywords and co-occurrences of keywords in a stream of documents, respectively. This network is used to rank and extract top-ranked emergent keywords that will also be used in summarization. Semantic graphs are among the most successful approaches that are broadly used for such tasks as keyword extraction and summarization \cite{litvak2008graph,yeh2008ispreadrank}, disambiguation \cite{sussna1993word}, and term similarities \cite{budanitsky2006evaluating,gabrilovich2007computing}. Perhaps, the most relevant work to our method is the SemanticRank approach  \cite{tsatsaronis2010semanticrank} that is a modification of a PageRank and HITS algorithms. However, to the best of our knowledge, in most existing works, only \emph{static} semantic graphs have been considered which does not resolve a problem of extracting emergent information in streaming settings. In a dynamic semantic graphs, nodes and edges can: (a) appear when previously have not been observed; (b) change their weights that represent the amount of keywords, and the connection strength between nodes, respectively, and (c) disappear when become obsolete after a certain time.  

A traditional way to detect emergent keywords is based on different quantities that directly depend on counting keyword frequencies. Examples include methods that rely on counting bursty keywords that suddenly appear at unusually high rates \cite{mathioudakis2010twittermonitor,goorha2010discovery} and a variety of methods that are based on the classical term frequency-inverse document
frequency (tf-idf) \cite{salton1988term} that evaluates an importance of a word with respect to a document in a corpus. In some methods, instead of considering the single term quantities, it is more appropriate to take into account these quantities only when the corresponding terms appear along with other highly frequent terms (see survey \cite{atefeh2015survey}). We propose to rank the importance of a keyword in a static semantic network using the eigenvector centrality \cite{bonacich2007some}, and then introduce the \emph{dynamic eigenvector centrality} to capture emergent keywords and summarize the trends. We also generalize the frequency and degree centralities with similar dynamic versions. However, in many cases, the eigenvector-based centrality is more illuminating as it has been observed and noticed multiple times that important keywords are likely to appear with other important keywords 
\cite{erkan2004lexrank,tsatsaronis2010semanticrank}. This concept is reflected in the eigenvector centrality ranking in which the importance of a keyword depends on the importance of co-occurring keywords. 

In the streaming data  setting, we consider a time line discretized into segments $t_i$, and introduce the dynamic eigenvector centrality ranking that takes into account the normalized eigenvector centralities on $P$ segments back from the current time segment $t$. We define the slope of an eigenvector centrality for a keyword $k$ at time segment $t$ as
\begin{equation}\label{eq:slp}
\slpec_k^{(t)} = \dfrac{\mathlarger{\sum}_{t_i \in \{ t,t-1,...,t-P\}} \left(t_i - \overline{T}\right)\left(\ec_k^{(t_i)} - \frac{1}{P}\sum_{i=0}^{P-1} \ec_k^{(t_i)} \right)}{\mathlarger{\sum}_{t_i \in \{ t,t-1,...,t-P\}} (t_i - \overline{T})^2},
\end{equation}
where $\overline{T} = P(P+1)/2$, and $\ec_k^{(t_i)}$ is the normalized eigenvector centrality of keyword $k$ at time segment $t_i$, i.e., $\slpec_k^{(t)}$ is a slope of a fitted linear regression model on normalized eigenvector centralities computed on $P$ time segments. Accordingly, we define the dynamic eigenvector centrality of keyword $k$ at time $t$ as
\begin{equation}\label{eq:dec}
\dc_k^{(t)} = \slpec_k^{(t)} \cdot \ec_k^{(t)}.
\end{equation}
Besides, a straightforward and easy computation (including a variety of methods to compute the eigenvector of a semantic graph \cite{vlsicad}), using the slope in centrality measure has several important advantages. First, it is not sensitive to missing values that could appear as a result if a keyword has not been used in a particular segment. Second, it is interpretable, which means that a domain expert user who will need to define a threshold to distinguish between emergent and regular keywords, can justify the choice.

Summarization of documents in each $t_i$ is performed by choosing a small subset of documents that contain top-ranked keywords. While this approach is not new, we demonstrate that extraction of documents that contain dynamic mutually emergent keywords provides much more relevant information than other comparable approaches. To evaluate the proposed method we compare it with the degree centrality, dynamic degree centrality in which a similar slope is computed for the degrees of nodes that correspond to keywords, non-dynamic eigenvector centrality, simple frequency ranking for all keywords, and dynamic frequency ranking (a similar slope is computed).

\section{Algorithms}
\label{sec:algorithms}
Introducing dynamic centrality emergent keyword extraction, and stream summarization, we also compare it to the dynamic degree centrality, keyword frequency count, and their corresponding non-dynamic versions. In all dynamic versions, similar to Equations (\ref{eq:slp}) and (\ref{eq:dec}), a regression slope is computed for the corresponding centrality indices, and then multiplied by them.
We process a data stream by discretizing it into time segments. The dynamic centralities are computed using the slopes fitted on last $P$ segments. At each time segment $t$, we maintain a semantic graph of keywords $G^{(t)} = (V^{(t)},E^{(t)})$, where $V^{(t)}$ is a set of nodes that correspond to keywords, and $E^{(t)}$ is a set of positive weighted edges that correspond to the number of co-occurences of two keywords in the same document, i.e., for keywords $i$ and $j$, there is an edge $ij\in E^{(t)}$ with weight $w^{(t)}_{ij}$ if $i$ and $j$ appear together in $w^{(t)}_{ij}$ documents. If a contribution of a document $d$ to $w^{(t)}_{ij}$ has been done $K$ time segments ago (where $K$ is a parameter determined by the application), the weight of $ij$ will be decreased at time $t+1$. Accordingly, $w^{(t)}_{ij}$ can be increased at time $t+1$ if $i$ and $j$ appear together again. As a result, an obsolete edge can be removed from the graph if its weight becomes 0. Obsolete nodes can also be removed if they become completely disconnected. In other words, $G^{(t)}$ will contain information of $\max(K, P)$ steps back. A degree of node $i$ at time $t$ is denoted by $d_i^{(t)}$. A frequency of a keyword (node) $i$ at time $t$ is denoted be $f_i^{(t)}$.

Below we describe six algorithms we experimented with to extract keywords.\\
\noindent $\blacktriangleright$ {\bf Degree centrality} All keywords $i\in V^{(t)}$ are ranked by normalized degrees $d_i^{(t)} / \max_i\{d_i^{(t)}\}$.\\ 
\noindent $\blacktriangleright$ {\bf Dynamic degree centrality} For each keyword $k\in V^{(t)}$, we consider $P$ values $d_k^{(t)}$, $d_k^{(t-1)}$, ... , $d_k^{(t-P-1)}$ to evaluate the slope $\slpdeg_k^{(t)}$ (similar to Equation (\ref{eq:slp})). The dynamic degree centrality is defined as 
\[
\ddeg_k^{(t)} =  \slpdeg_k^{(t)} \cdot d_k^{(t)}.
\]
\noindent $\blacktriangleright$ {\bf Frequency centrality} All keywords $i\in V^{(t)}$ are ranked by their frequencies $f_i^{(t)} / \max_i\{f_i^{(t)}\}$.\\ 
\noindent $\blacktriangleright$ {\bf Dynamic frequency centrality} For each keyword $k\in V^{(t)}$, we consider $P$ values $f_k^{(t)}$, $f_k^{(t-1)}$, ... , $f_k^{(t-P-1)}$ to evaluate the slope $\slpfreq_k^{(t)}$ (similar to Equation (\ref{eq:slp})). The dynamic degree centrality is defined as 
\[
\dfreq_k^{(t)} =  \slpfreq_k^{(t)} \cdot f_k^{(t)}.
\]
\noindent $\blacktriangleright$ {\bf Eigenvector centrality} All keywords $i\in V^{(t)}$ are ranked by the entries of the eigenvector $x$ in solving $A^{(t)}x = \lambda x$, where $A^{(t)}$ is a weighted adjacency matrix of $G^{(t)}$, $x$ is the eigenvector that correspond to the largest eigenvalue of $A^{(t)}$. The normalized centrality index for a keyword $i$ is then defined as $\ec_i^{(t)} = x_i / \max_i\{x_i\}$.\\
\noindent  $\blacktriangleright$ {\bf Dynamic eigenvector centrality} See Equation (\ref{eq:dec}).

In all cases, a positive slope indicates an increase in significance of the keyword while a negative slope shows an opposite trend. Hence, when multiplying by the slopes, the less important words can gradually be removed (if we set up a threshold of importance or use an appropriate insignificant outlier detection method), and there is a boost in value of keywords with high slope. By picking high value keywords, we pick the trending keywords. While optimizing the running time is not the goal of this paper, it is clear that the most  computationally intensive part is a computation of the eigenvector which a well studied topic \cite{hernandez2005slepc,brouwer2011spectra,vlsicad}. 

The pseudocode for computing dynamic eigenvector centralities is summarized in Algorithm \ref{fig:pseudocode:dec}. Two input parameters are the graph $G^{(t-1)}$ from step $t-1$ that will be updated with the current step data $D^{(t)}$. The $D^{(t)}$ is preprocessed with the following steps that are relevant to Twitter data: (1) convert text to lowercase, (2) remove special characters, (3) lemmatize words, (4) remove html tags, and (5) remove stop words. 

\begin{figure}[h]
\begin{algorithmic}[1]
\Procedure{DEC}{$G^{(t-1)}$, $D^{(t)}$}
\State Initialize $G^{(t)}$ with $G^{(t-1)}$
\State Update $V^{(t)}$ and $E^{(t)}$ by decreasing $w_{ij}^{(t)}$ and dropping   obsolete edges and nodes
\State Update $V^{(t)}$ and $E^{(t)}$ by adding new edges and keywords from $D^{(t)}$
\State Compute $\ec_i^{(t)}$ for $G^{(t)}$
\State Rank all keywords by $\dc_i^{(t)}$ for $G^{(t)}$
\State $S \leftarrow$ top-ranked $K$ keywords
\State \Return {$S$ and $G^{(t)}$}
\EndProcedure
\end{algorithmic}
\caption{Algorithm for computing $\dc_{\cdot}^{(t)}$}\label{fig:pseudocode:dec}
\end{figure}

The extracted top-ranked emergent keywords are used to summarize the data stream. The summarization is done by finding a small set of documents that cover the entire set of top keywords (see $S$ in line 7 of Algorithm \ref{fig:pseudocode:dec}). We formulate the minimum size set cover problem, where $S$ is the set to be covered, and documents are the subsets of keywords that participate in the covering. It is interesting to mention that finding the real minimum number of documents that cover $S$ may not be informative enough because the information can be too compressed. Thus, we decided to use the greedy set cover algorithm \cite{chvatal1979greedy} that is fast enough but does not compress the summarization too much because of the obvious reasons of poor approximation ratio. In this setting, every document $i$ is associated with a weight 
\[
c_i = \sum_{k\in S} \tf (k,i),
\]
where $\tf (k,i)$ is a frequency of keyword $k\in S$ in document $i$. The weight of the rest of the keywords is zero. In the greedy algorithm, we repeatedly select document $i$ that minimizes $c_i / |S \setminus C|$, where $C$ is the list of already covered (in previous steps of a greedy algorithm) keywords in $S$. There could be a situation where the emergent keyword may not be present in the documents of that time segment. In such cases, the algorithm is run until it covers top keywords in that hour. The selected documents represent a summary based on the emergent keywords.




\section{Experiments and discussion}

How good is our proposed dynamic eigenvector centrality measure in detecting emergent keywords and summarization? In this section we evaluate our method with Twitter data collected for two public safety events: the 2013 Boston Marathon attacks and the 2015 Baltimore protests (further details for each events are provided below). Both events are characterized by high volumes of Twitter activity and rapidly changing language and emergent terms used to describe unfolding events, which makes them ideal test cases to evaluate our method's ability to detect emergent keywords and summarize the data stream. 

For each event, we purchased archived tweets from Gnip, a company that provides access to the full archive of public Twitter data. We used broad search terms to collect tweets in order to create noisy data streams that cover both related and unrelated events. Our data set for Boston covers seven days from April 15--21, 2013, and was collected with the following search terms: boston, marathon, bomb, blast, explosion, watertown, mit, mitshooting. For Baltimore, we collected tweets for a 15 day period from April 18--May 2, 2015, using the following search terms: joseph kent, freddie gray, eric garner, ferguson, curfew, police, riot, protests, loot, looting, \#purge, \#baltimore, \#baltimoreriots, \#baltimoreuprising, \#freddiegray, \#josephkent, \#blacklivesmatter, \#onebaltimore, rioter, charge, charged, murder, homicide, mosby, corporal, \#mayday, justice, \#blackspring, \#freddiegray's, cops, unjustified, spinal, broken spine, arrested, thugs, thug, \#marilynmosby, \#wakeupamerica, freddie, racist, racism, \#baltimoreprotest, propaganda, officers, knife. Table~\ref{tab:data} provides an overview of the data used for the experiments. A logical OR expression was used to filter the terms in both cases, i.e., for example, by keeping term \emph{boston}, we obtained all tweets related to the city, and not necessary to the bombing event.
\begin{table}
\caption{Overview of data used for experiments\label{tab:data}}
\begin{center}
\begin{tabular}{lcc} 
\hline
                 & Boston (2013)         & Baltimore (2015 \\ \hline
Time period      & 8pm, April 14 --      & 8pm, April 17 --\\
                 & 23.59pm, April 21     & 23:59pm, May 2 \\
Number of tweets & 20,385,957            & 19,763,762 \\ \hline
\end{tabular}
\end{center}
\end{table}

For each of the two public safety events, we coded major occurrences and changes in events from information published by news outlets. We use the timing of these events as ground truth against which we compare the algorithms discussed in Section~\ref{sec:algorithms}. Before applying each algorithm to the tweet texts (i.e., the maximum 140 character long texts that users posted publicly), we followed standard pre-processing procedures, including the removal of stop words, numbers, URLs, and all tweets in a language other than English. We further grouped tweets into one-hour time segments. 

To calculate the dynamic versions of each measure, we set $P=5$. That is, we weighted each measure at time $t$ with the slope of a linear regression fitted to the last five time segments. 

Using actual events as ground truth, we conducted two types of experiments. First, as a proof-of-concept, we use time series plots to inspect how well different measures of keyword importance detect emergent ground-truth events. Second, we show summaries of key emergent events created by the minimum size set cover algorithm for the Boston event.

\subsection{Event 1: Boston Marathon bombing}
In our first set of experiments, we look at changes in events surrounding the 2013 Boston Marathon attacks, which occurred when two bombs were detonated close to the marathon finishing line on April 15, killing three and injuring more than 260 people. The detonation of the two bombs was followed by a four-day manhunt for the terrorists, ending on the evening of April 19 after one attacker was killed in a gun battle with police and the second attacker, Dzhokhar Tsarnaev, was found hiding in a boat in the back yard of a house in Boston's Watertown neighborhood.

We identified six key events from time lines published by two news outlets \cite{CNNBoston,BostonCom}: (1) the detonation of the two bombs at 2:49pm on April 15, (2) the explosion of a fertilizer company at 7:50pm on April 17 in Texas, which was unrelated to the events in Boston, but was briefly thought to be another terrorist attack, (3) the publication of surveillance photos and videos of the two suspects at 5pm on April 18, (4) the death of a MIT police officer at 10:30pm on April 18, (5) the official release of Tsarnaev's name and photo at 7am on April 19, and (6) his capture by police while hiding in a boat in Watertown at around 8:45pm on April 19.

\subsubsection{Experiment 1: Estimated keyword importance versus ground truth}
\begin{figure*}
\begin{center}
\includegraphics[width=\textwidth]{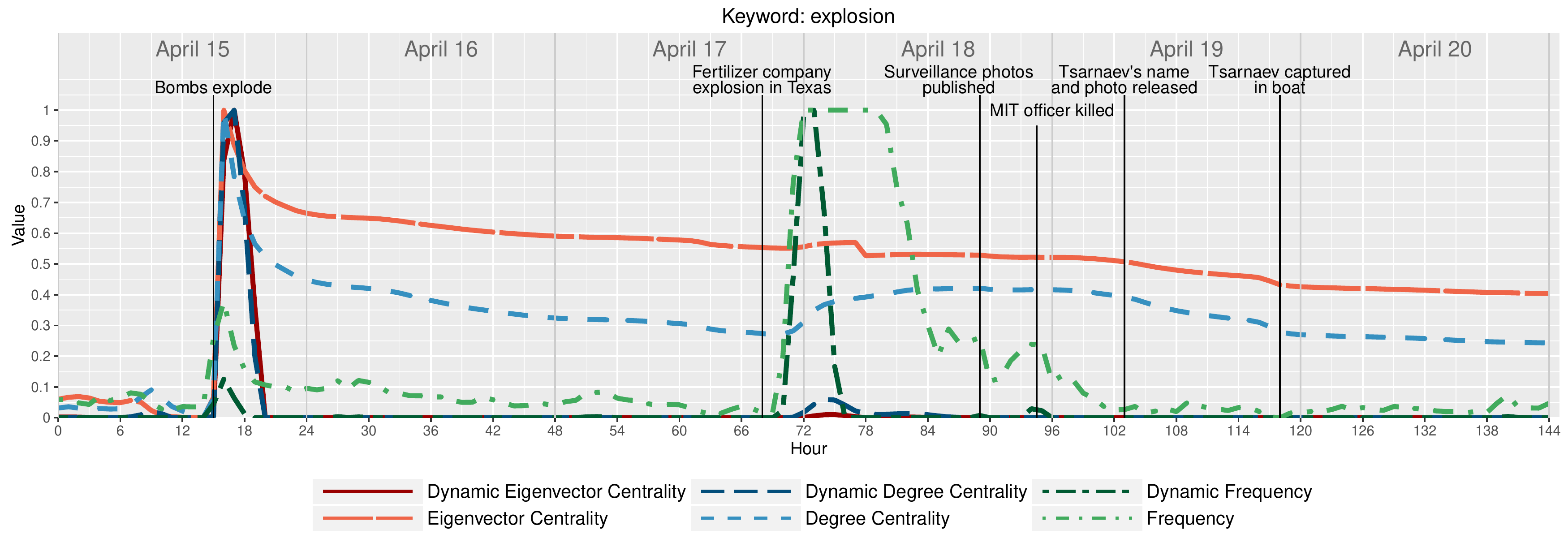}
\caption{Timeline of actual events during Boston Marathon attacks together with importance measures for keyword ``explosion''. \label{fig:Boston_timeline_explosion}}
\end{center}
\end{figure*}

\begin{figure*}
\begin{center}
\includegraphics[width=\textwidth]{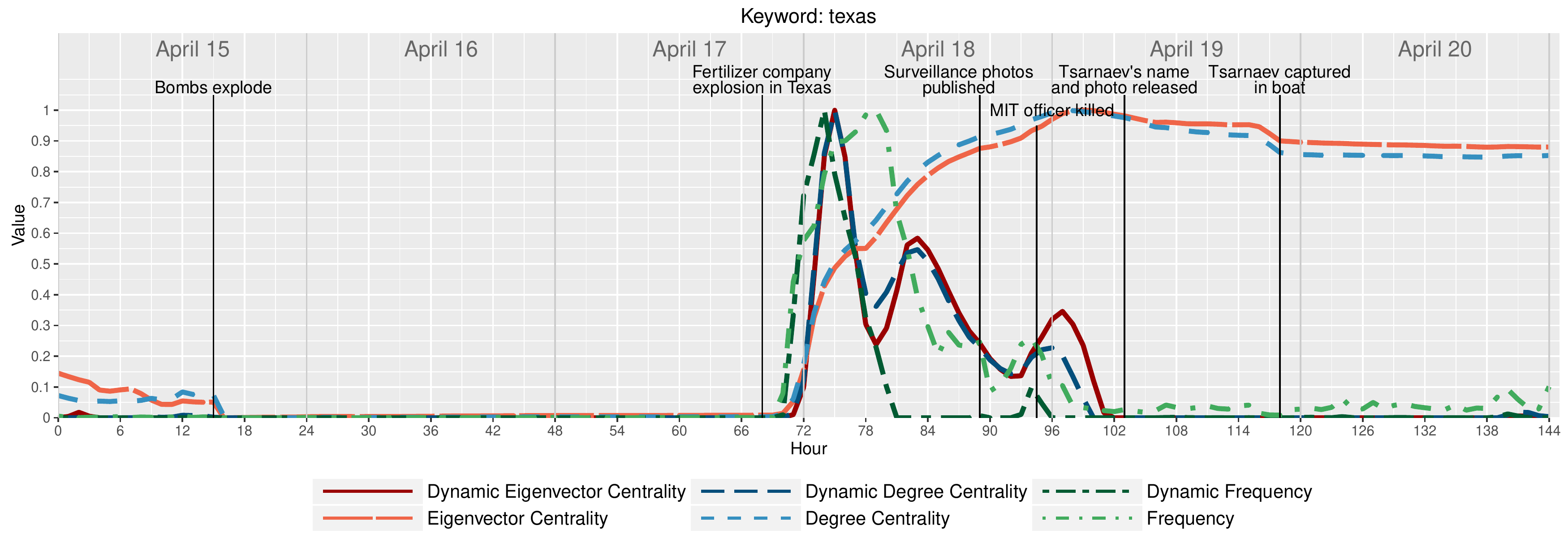}
\caption{Timeline of actual events during Boston Marathon attacks together with importance measures for keyword ``texas"'. \label{fig:Boston_timeline_texas}}
\end{center}
\end{figure*}

\begin{figure*}
\begin{center}
\includegraphics[width=\textwidth]{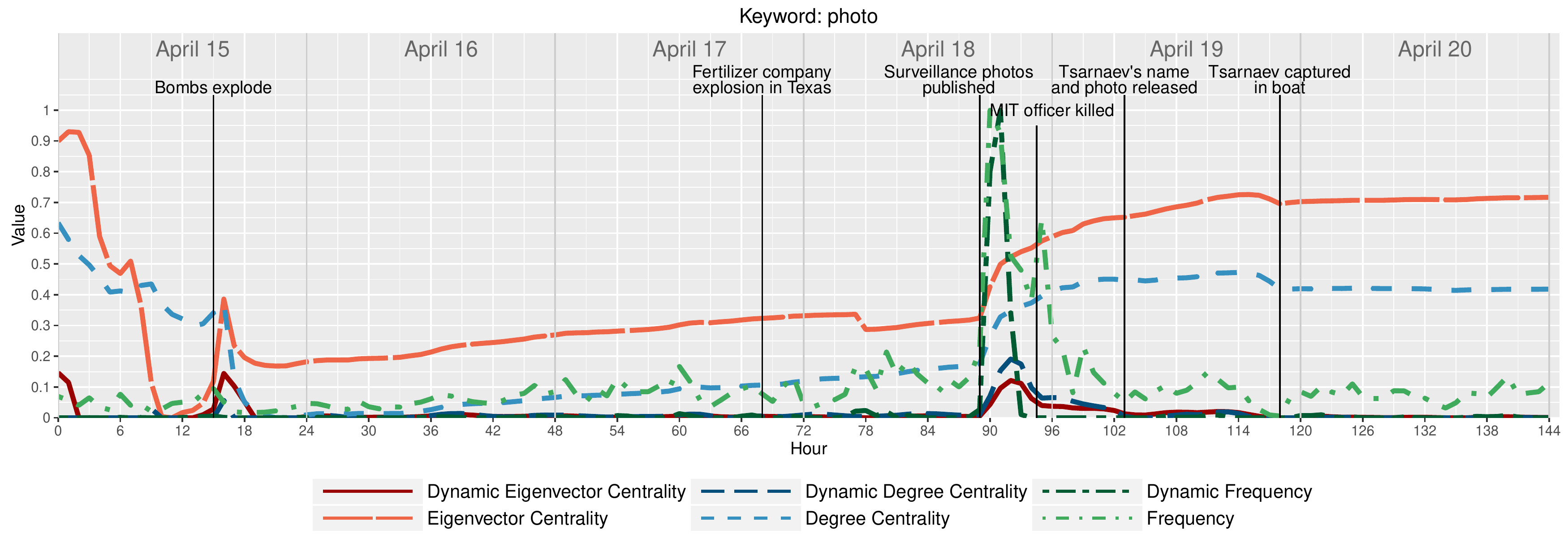}
\caption{Timeline of actual events during Boston Marathon attacks together with importance measures for keyword ``photo"'. \label{fig:Boston_timeline_photo}}
\end{center}
\end{figure*}

\begin{figure*}
\begin{center}
\includegraphics[width=\textwidth]{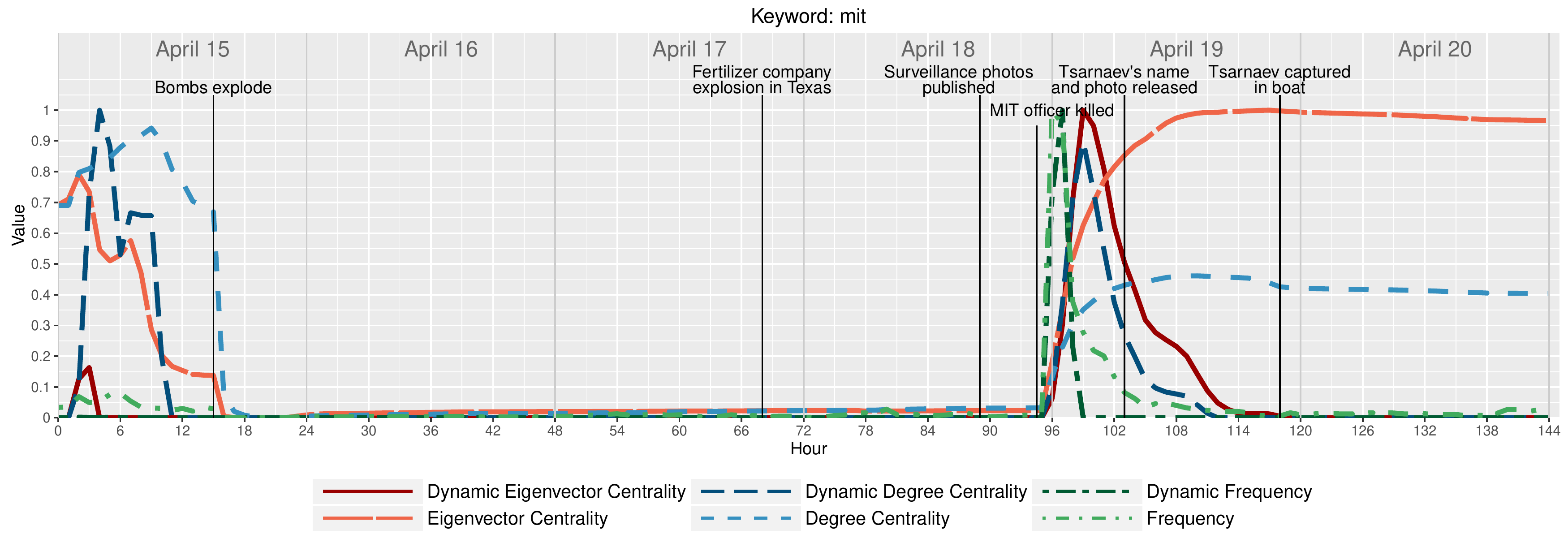}
\caption{Timeline of actual events during Boston Marathon attacks together with importance measures for keyword ``mit. \label{fig:Boston_timeline_mit}}
\end{center}
\end{figure*}

\begin{figure*}
\begin{center}
\includegraphics[width=\textwidth]{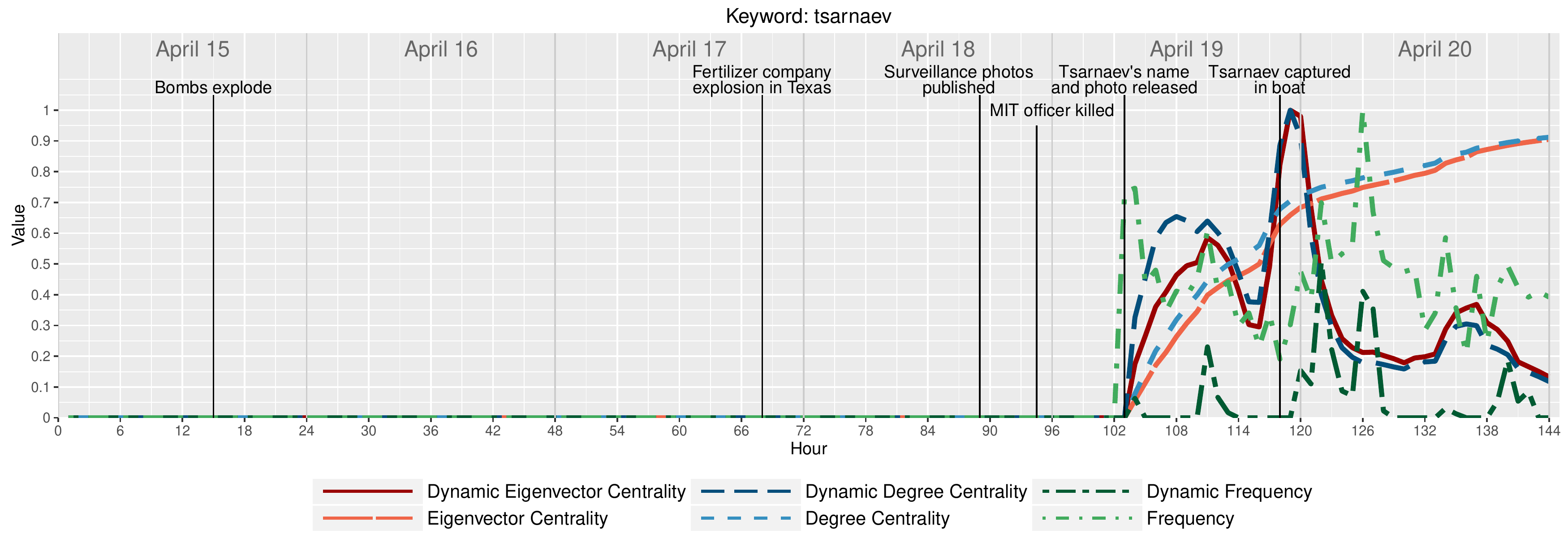}
\caption{Timeline of actual events during Boston Marathon attacks together with importance measures for keyword ``tsarnaev''. \label{fig:Boston_timeline_tsarnaev}}
\end{center}
\end{figure*}

In our first experiment, we compare the six importance measures to the ground truth events. To allow for a direct comparison between the measures, we first replace negative values in the dynamic measures to zero and then normalize each measure to $[0,1]$. Figures~\ref{fig:Boston_timeline_explosion} to \ref{fig:Boston_timeline_tsarnaev} show the ground-truth time lines together with the importance measures for six selected keywords most closely related to the actual events: ``explosion'', ``texas'', ``photo'', ``mit'', ``tsarnaev'', and ``boat''. 

The following conclusions can be drawn from the six figures:
\begin{itemize}
\item As a proof-of-concept, the figures show that keyword-based dynamic centrality measures are extremely capable in detecting emergent events. In all cases, large spikes in the dynamic measures closely follow corresponding ground truth events. 
\item The dynamic measures are superior to their static versions when it comes to labeling keywords as emergent. For example, both dynamic eigenvector centrality and dynamic degree centrality for keyword ``explosion'' sharply increase shortly after the explosion of the two bombs, but then---and in contrast to their static counterparts---decrease keyword importance to zero. 
\end{itemize}

\begin{figure*}
\begin{center}
\includegraphics[width=\textwidth]{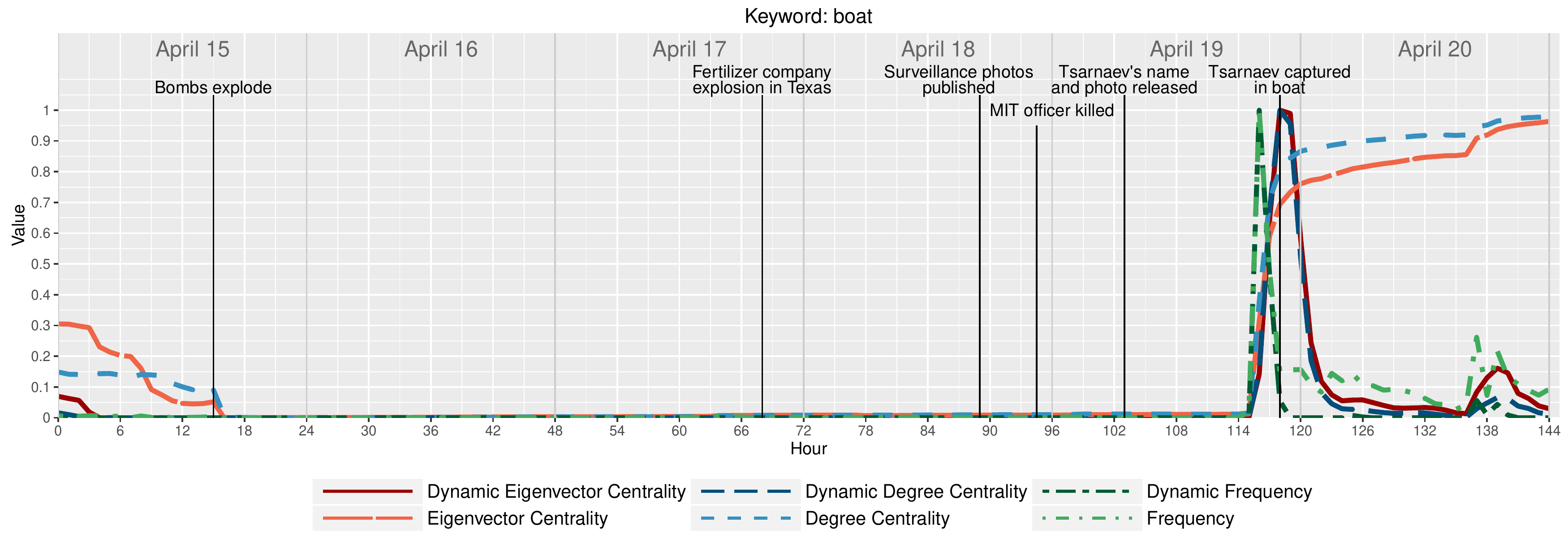}
\caption{Timeline of actual events during Boston Marathon attacks together with importance measures for keyword ``boat''. \label{fig:Boston_timeline_boat}}
\end{center}
\end{figure*}

\subsubsection{Experiment 2: Summary of emerging events for Boston Marathon bombing}
Having established that dynamic eigenvector centrality is a suitable measure to detect emergent events, we use it to generate a summary of the data stream. For each hour, we select the top 20 ranked keywords and then apply the greedy set cover algorithm \cite{chvatal1979greedy} discussed in Section~\ref{sec:algorithms} to find the smallest number of tweets that cover the 20 keywords. The result is a set of documents that represent a summary of the stream based on the  emergent keywords.

Tables~\ref{tab:summaries_bombing1}, \ref{tab:summaries_bombing2}, and \ref{tab:summaries_bombing3} show the top 20 keywords and summaries for the three hours before, during, and after the explosion of the bombs during the Boston Marathon at 2:49pm. The hour before the attacks is characterized by conversations around the marathon winners Lelisa Desisa from Ethiopia and Rita Jeptoo from Kenya, with their names, countries, and words such as ``win'', ``won'', and ``winner'' being among the top keywords. The keywords and summaries for the next hour from 2:00-3:00pm, which covers the attacks at 2:49pm, are still dominated by marathon-related conversations, but now include the keyword ``explosion'' and a corresponding tweet (``Explosion! [url]''). In the following hour, the keywords and summaries have completely shifted to language describing the emergent event (``breaking'', ``news'', ``explosion'', ``bomb''), its location (``Boston'', ``finish'', ``line''), and offering condolence (``prayer'', ``praying'', ``thought'').


\begin{table}
\caption{Top 20 keywords and tweet summaries based on dynamic eigenvector centrality during the three hours before, during and after the explosions at 2:49pm. \label{tab:summaries_bombing1}}
\begin{tabular}{p{\textwidth}}
\hline \hline
\multicolumn{1}{c}{\textsc{Hour: 1:00--2:00pm}} \\
\hline
\textbf{Top 20 keywords:}  boston, win, desisa, run, men, lelisa, jeptoo, woman, won, ethiopia, mile, rita, finish, race, kenya, watching, winner, ha, half, kenyan \\
\textbf{Tweet summaries:} "kenyans winning a marathon is hardly news. havent they done it for the last 3 decades or something?", "just realized that once finals week rolls around I will have zero finals and already ran in the half marathon...I fear for my health", "MT @brendan207: Looking for marathon Monday lunch?  @boloco has u covered!  Having a delicious lunch @BolocoCongress right now! \#PatriotsDay", "Our students competed in an exciting \#Math Multiplication Marathon. Congratulations to our grade level winners! [url]", "Policeman letting his son wear his hat while watching the marathon together...too cute not to share [url]", "Museveni ICC blast in Kenya was not a surprise to West [url] via @dailymonitor", "@BostonMarathon is America's oldest marathon, the yardstick by which these other foot races are measured: [url]", "Does Joe have a goal time for the marathon or does he just want to finish? \#RunJoeyRun", "Running your own biz is like running a marathon: -- Start to 7 miles: Find your pace; SET YOUR GOALS; warm up the... [url]", "help out a good cause please [url] \#charity \#ethiopia \#buildinghouses \#donate \#HabitatForHumanity \#marathon \#coffee \#buns", "Rita Jeptoo averaged 5:35 pace for her marathon \#CrushedIt", "Well Jasmine got one of her wishes today, an Ethiopian won her hometown marathon.", "'When a woman asks you to guess her age.... is like deciding which wire to cut to. difuse the a bomb'.", "Four 'in TA base toy-car bomb plot': A court hears details of how four British men talked about bombing a Terr... [url]", "Pissed I couldn't go to the marathon to see my sister run :( \#GoodLuck", "Imagine what he'll do for the third: Lelisa Desisa wins \#BostonMarathon in just his second marathon ever. [url]", "Africans prevail in Boston Marathon [url]" \\
\hline
\hline
\end{tabular}
\end{table}

\begin{table}
\caption{Top 20 keywords and tweet summaries based on dynamic eigenvector centrality during the three hours before, during and after the explosions at 2:49pm. \label{tab:summaries_bombing2}}
\begin{tabular}{p{\textwidth}}
\hline \hline
\multicolumn{1}{c}{\textsc{Hour: 2:00--3:00pm}} \\
\hline
\textbf{Top 20 keywords:} boston, win, desisa, lelisa, finish, run, rt, jeptoo, men, ethiopia, won, rita, woman, mile, time, explosion, ha, kenya, winner, race \\
\textbf{Tweet summaries:} "corrib road race  marathon monday", "Friendship Blast Contest Winner's Photoshoot!! [url]", "Too much rain in Kenya, i will resume exercise tomorrow ! 'Marathon '", "Officially on the last Harry Potter film... This marathon has shown me I have emotions I didn't even know existed \#hpmarathontroops", "Explosion! [url]", "First time I've worked Marathon Monday in the 21st Century. No sir, I don't like it.", "My sister just asked me to do a 26 mile marathon with her in September. Does she even know who I am", "Had a blast working on this beauty!  Stay tuned for the video!  Like us on FB Art Of A Woman [url]", "Rita Jespoo is proof that mat leave is awesome [url] \#Bostonmarathon", "He won marathon Monday [url]", "Marathon and other victories by our athletes in many world cities are purely organic, natural. No doping... \#Oromo \#Oromia \#Ethiopia", "We had a blast at Homecoming on the Hill 2013! Here's a great photo of all those who participated in the Men's... [url]", "RT @andrewbensonf1: BBC News - Car blast in Bahrain heightens F1 security concerns [url]", "Surround yourself w/ppl who push you. I have no desire to run a full marathon but I am inspired to push myself beyond what I think I can do.", "The fact that my aunts just got VIP passes for the marathon, see you at the finish line @DColl15, legit", "@Yusufdido @robjillo  Lelisa means 'someone who desires' in Afan Oromo in Oromia (and he desired the Marathon and he got it)", "Rita Jeptoo wins in 2:26:25 \#bostonmaraton”. Again this would be maybe what I could do a half marathon time in. Crazy fast", "@kpfallon obvs. I mean, Lelisa Desisa and I have so many similarities, winning only our second marathon will be just one.", "I posted 7 photos on Facebook in the album '2013 Boston Marathon' [url]" \\
\hline
\hline
\end{tabular}
\end{table}

\begin{table}
\caption{Top 20 keywords and tweet summaries based on dynamic eigenvector centrality during the three hours before, during and after the explosions at 2:49pm. \label{tab:summaries_bombing3}}
\begin{tabular}{p{\textwidth}}
\hline \hline
\multicolumn{1}{c}{\textsc{Hour: 3:00--4:00pm}} \\
\hline
\textbf{Top 20 keywords:} explosion, finish, line, boston, prayer, people, thought, news, injured, rt, reported, praying, breaking, hope, happened, bomb, affected, report, bombing, safe \\
\textbf{Tweet summaries:} "Thanking god my aunts and uncle are safe at the marathon. \#prayforboston", "What do you get out of bombing a marathon someone please tell me. \#PrayForBoston", "\#Bostonmarathon Early report in the times: [url]", "The marathon this year was dedicated to everyone affected by newtown. It's sick someone could do that.", "@murneybhoy @saintturbo if I never saw another smoke bomb, gimp mask   upside down protest banner again @ our stadium i would rejoice!", "Omg and my daddy was gonna run in that marathon!!! I would have died if anything happened to him... Daddy's girl", "I hope my friend Irene is okay at the Marathon", "My aunt, who is 45 with MS is running in the marathon and it's breaking my heart that this is happening.", "\#prayersforboston praying for all who's at the marathon.", "@GeorgeSandeman Casualties reported now [url] F***ing hell :(", "RT @DaveWedge: ABC reporting 2 dead at marathon; dozens hurt; source tells me FBI counterterrorism team from NYC en route to \#BostonMarathon", "Pray for all those who are injured at the marathon and pray that the scum who did it rots in hell", "@HannahLuke23 heard the news about the marathon, are you alright?", "Thoughts go out to the victims in the Bodton Marathon attack.", "Terrorist attack at a marathon people raising money for good causes what's wrong with the world.", "Prayers for those at home and involved with the marathon.", "Boston...damn :(", "Its a directed blast both went off with the blast directed at the finsh line", "@ITK\_AGENT\_VIGO Looks like someone has set off a semtex Catherine wheel at the finish of the marathon.God will help the Yanks,he always does", "@GreenDayTilIDie lot of bad injuries. They saying more than one explosion" \\
\hline \hline
\end{tabular}
\end{table}

Tables~\ref{tab:summaries_boat1} and  \ref{tab:summaries_boat2} provide a second example of keyword selection and stream summary for the two hours covering the capture of Tsarnaev. The tweet summary clearly captures the event, including tweets such as ``'It's over' - CNN \#boston'' or ``They finally got that boy. I know Boston feelin good now.''

\begin{table}
\caption{Top 20 keywords and tweet summaries based on dynamic eigenvector centrality during the three hours before, during and after Tsarnaev's capture at 8:45pm. \label{tab:summaries_boat1}}
\begin{tabular}{p{\textwidth}}
\hline \hline
\multicolumn{1}{c}{\textsc{Hour: 8:00--9:00pm}} \\
\hline
\textbf{Top 20 keywords:} suspect, police, bombing, rt, news, watertown, ha, custody, bomber, breaking, manhunt, cnn, fbi, guy, area, officer, live, shot, terror, scanner \\
\textbf{Tweet summaries:} "Congratulations going around on the Boston PD scanner. And well deserved.", "i wonder if the hunger strikers in solitary confinement at guantanamo bay know about the alleged terror attacks in boston... prolly not aye", "Sure is a lot of shooting going on in Boston-Bin Laden did not get shot at that much-strangethings going on in Boston-I don't buy into it.", "@Sploops Go to WCVB TV Boston's live feed. They have been very good. Also WBUR radio.", "Do they have a suicide prevention officer there to talk to him? I hope so. Boston", "'@OnlyInBOS: Tomorrow is 420, which the Boston area really, really needs after this week.' My point exactly smh", "There's boys running round East Belfast have done worse shit than this guy, don't see Belfast being locked down anytime soon \#boston", "Good job, Boston PD, FBI et al. Deep breaths, deep breaths, everybody.", "'It's over' - CNN \#boston", "They found the 2nd dude that bomb Boston /.\ \#manhunt", "[ALERT] Boston Mayor Tom Menino says on twitter 'we got him' | Reuters \#Breaking", "\#blowthatboatup \#boston \#bostonmarathon \#bombers \#bp \#america \#staystrong [url]", "They have him in custody!!! \#BOSTON", "@KF***INGP after all Boston has been through the past few days... Give them a break Kenny! Hahaha", "YES!!!!! Take that stupid terrorist, you where told that we will find you and we did!!! \#watertown", "@FitzTheReporter Local news in Boston says so.", "RT @PrincessProbz: god bless america, god bless boston,   god bless all the victims   their family members. you are all in our prayers.", "Thoughts and Prayers to: the victims of the Boston Bombings and to the victims of irresponsible gun laws and policies in the U.S", "@Boston\_Police amazing job", "The Boston suspect 'captured' was found ALIVE hiding in a waste container! No mames" \\
\hline
\hline
\end{tabular}
\end{table}

\begin{table}
\caption{Top 20 keywords and tweet summaries based on dynamic eigenvector centrality during the three hours before, during and after Tsarnaev's capture at 8:45pm. \label{tab:summaries_boat2}}
\begin{tabular}{p{\textwidth}}
\hline \hline
\multicolumn{1}{c}{\textsc{Hour: 9:00--10:00pm}} \\
\hline
\textbf{Top 20 keywords:} suspect, police, bombing, rt, custody, watertown, news, ha, bomber, breaking, manhunt, fbi, officer, guy, terror, area, cnn, job, bostonstrong, good\\
\textbf{Tweet summaries:}  "They finally got that boy. I know Boston feelin good now.", "Tales at \#Boston didn't have to invade \#Iraq \#BostonStrong \#USA \#merica", "So many props and love for all the law enforcement in \#Boston, amazing job.", "MT @josh\_levin CNN's Susan Candiotti: 'Streets are empty. It's eerie. It's as though [pause, conjure metaphor] a bomb had dropped somewhere'", "Soldiers in Boston area streets are Massachusetts National Guard who proudly trace their roots to the Minutemen.", "Boston bombs: Obama lulled America into false confidence over terror threat   [url]", "Something about chanting USA doesn't seem appropriate. This guy was an American student, but I sure am happy they caught the SOB. \#Boston", "@WillSasso A lot of those officers were national, not from Boston", "After cheering subsides in \#boston, expect some serious questions to be asked about earlier investigation of \#bombingsuspect. \#fbi", "Final thought:  I feel like this Boston manhunt prevented us from really being able to talk about that new Daft Punk single.", "Releasing photos a risk, but pivotal in breaking \#Boston case. [url] \#BostonManhunt", "Glad to see they caught the Boston Bomber. Now I can go back to posting and tweeting about nonsense and less important things.", "My thoughts with the families who had lost their loved ones. Justice tonight in boston has been served", "Boston PD press conference coming up at 9:30. Tune in to CBS 13 News", "Cheers @KellyMacFarland who's been on lockdown right next to the shootout in Watertown. Go catch one of her shows and send her a drink!!!", "Dzhokhar Tsarnaev is lucky he is in Boston and not Los Angeles. LAPD could take some lessons from BPD! So glad that he is in custody alive.", "RT @msdebbieallen: My Heart goes out to our families in Boston and Texas. President Obama said it well ... [url]", "'@gherpich99: This Boston bombing story is crazy!'", "Bravo, Boston Police Department!! Bravo!!! \#bostonpolice", "WashPo give the skinny on the fate of the Boston bomb suspects [url]" \\
\hline \hline
\end{tabular}
\end{table}

\subsection{Event 2: Baltimore protests}
Our next set of experiments is based on tweets posted during the 2015 Baltimore protests. The protests were in response to the death of Freddie Gray, an African American resident of Baltimore, who died while in police custody. Gray's arrest and news of his death caused a series of protests, civil unrest and riots, and resulted in a city wide curfew that came into effect in the evening of April 28. 

\subsubsection{Experiment 1: Timeline of actual events and emergent keywords for Baltimore protests}
\begin{figure*}
\begin{center}
\includegraphics[width=\textwidth]{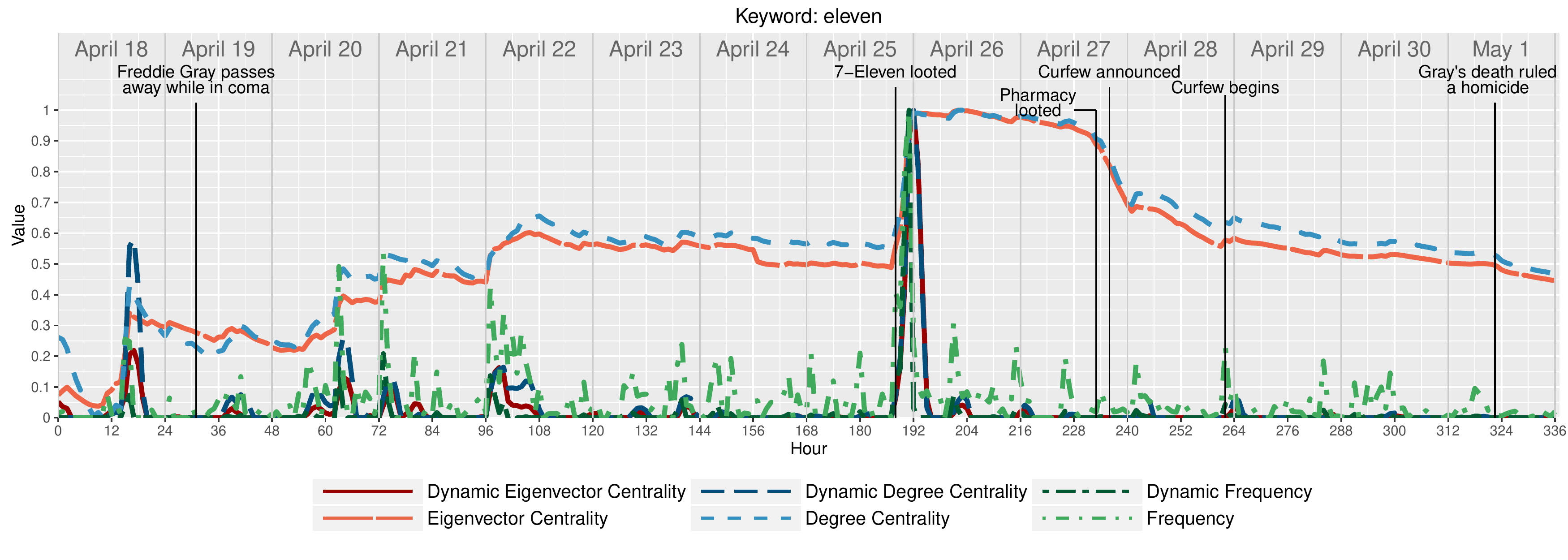}
\caption{Timeline of actual events during Baltimore protests together with importance measures for keyword ``eleven''. \label{fig:Baltimore_timeline_eleven}}
\end{center}
\end{figure*}

\begin{figure*}
\begin{center}
\includegraphics[width=\textwidth]{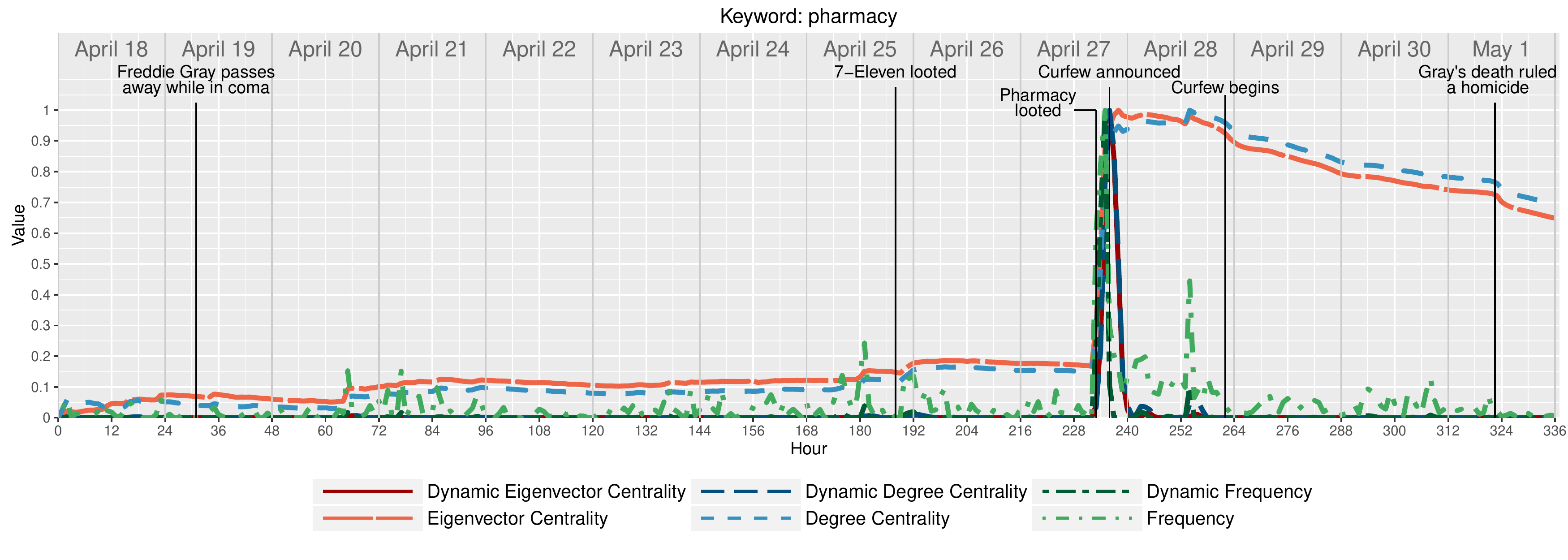}
\caption{Timeline of actual events during Baltimore protests together with importance measures for keyword ``pharmacy''. \label{fig:Baltimore_timeline_pharmacy}}
\end{center}
\end{figure*}

\begin{figure*}
\begin{center}
\includegraphics[width=\textwidth]{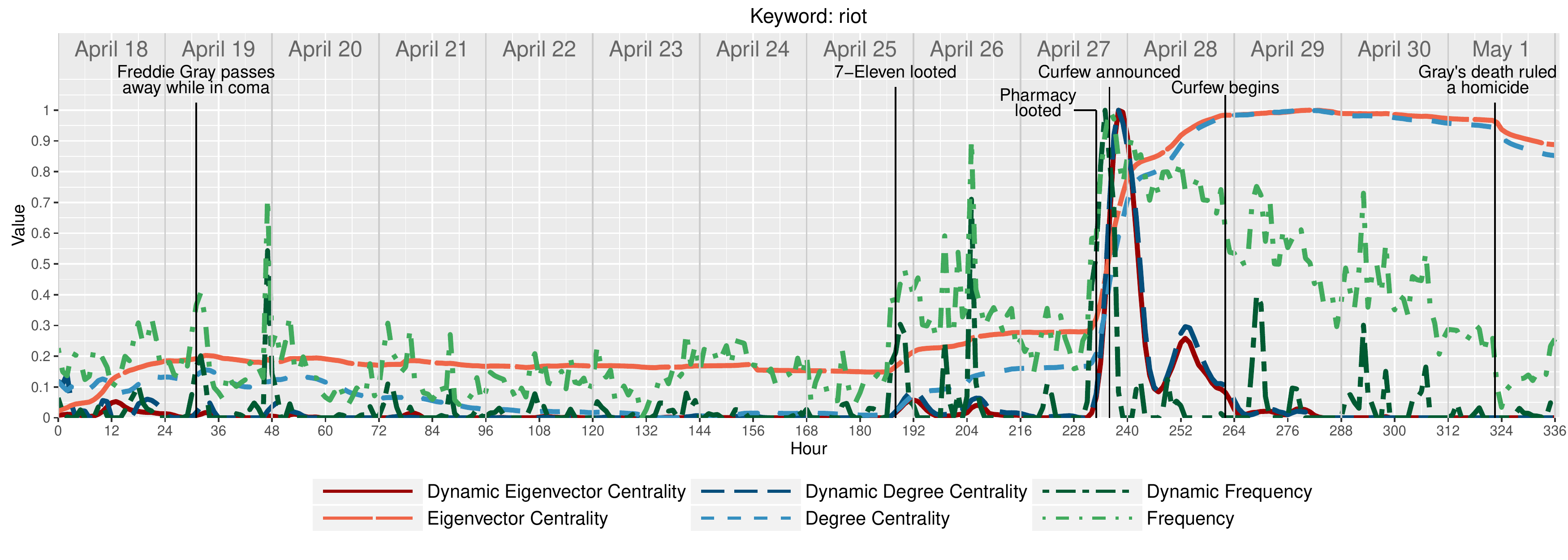}
\caption{Timeline of actual events during Baltimore protests together with importance measures for keyword ``riot''. \label{fig:Baltimore_timeline_riot}}
\end{center}
\end{figure*}

\begin{figure*}
\begin{center}
\includegraphics[width=\textwidth]{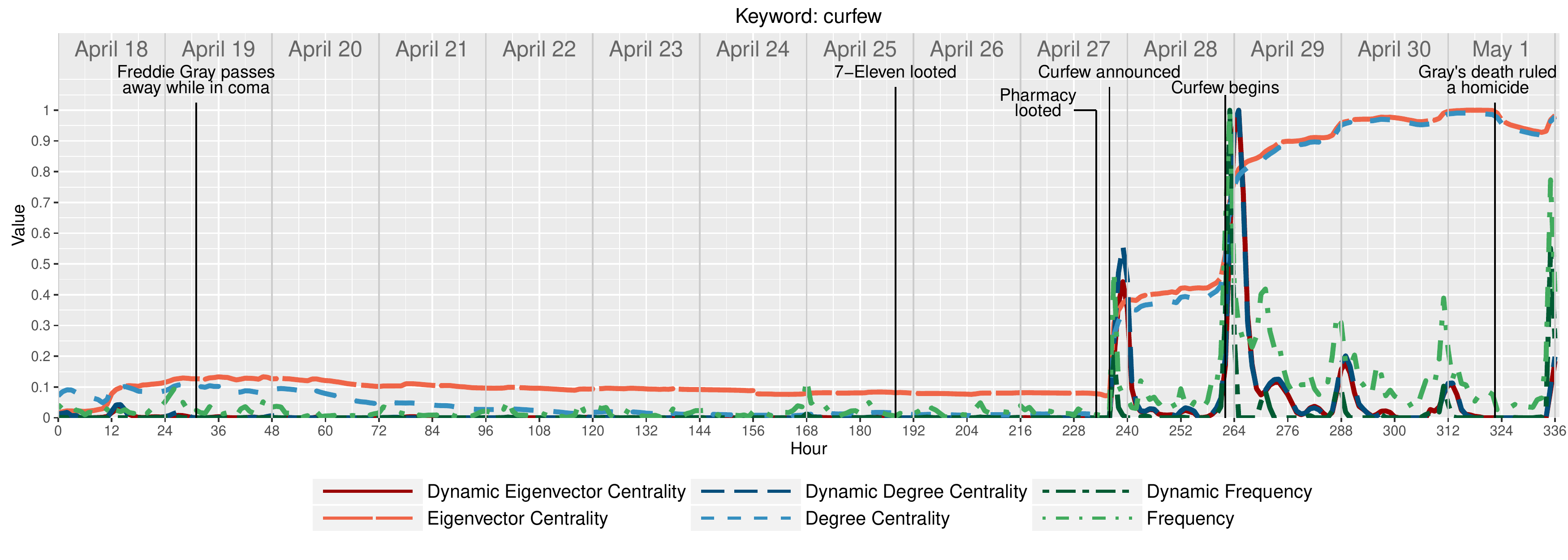}
\caption{Timeline of actual events during Baltimore protests together with importance measures for keyword ``curfew''. \label{fig:Baltimore_timeline_curfew}}
\end{center}
\end{figure*}

\begin{figure*}
\begin{center}
\includegraphics[width=\textwidth]{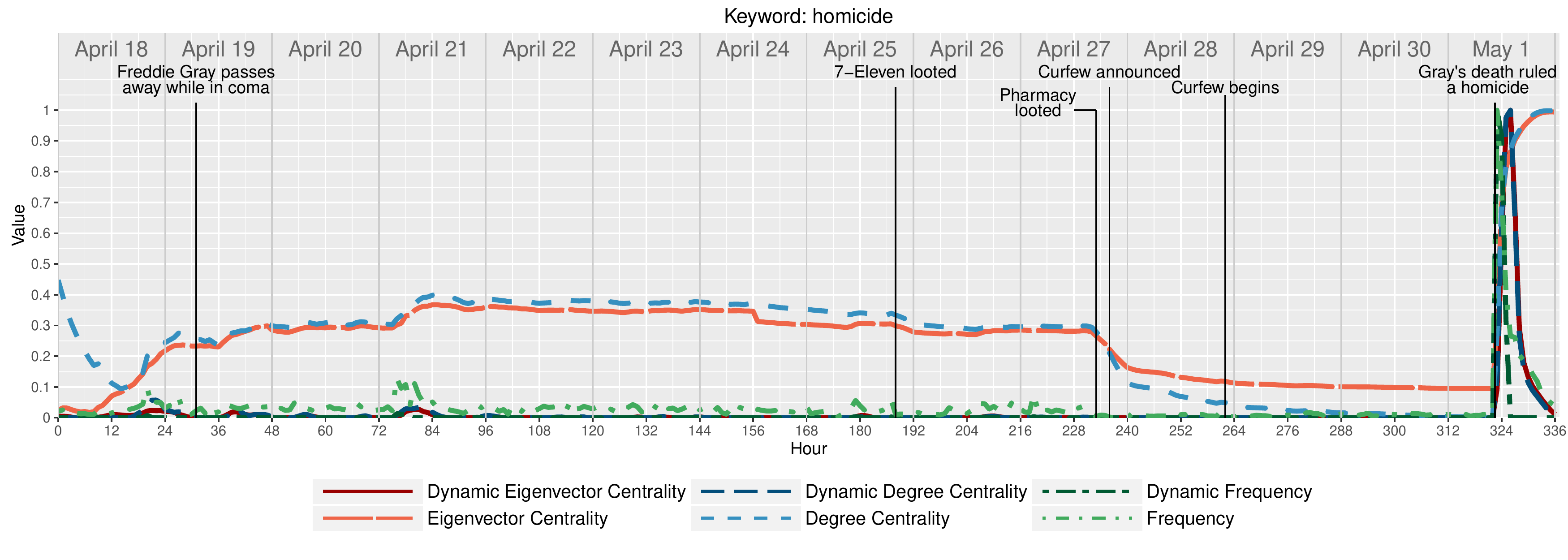}
\caption{Timeline of actual events during Baltimore protests together with importance measures for keyword ``homicide''. \label{fig:Baltimore_timeline_homicide}}
\end{center}
\end{figure*}

We identified six key events that we plot together with the six importance measures in Figures~\ref{fig:Baltimore_timeline_eleven} to \ref{fig:Baltimore_timeline_homicide}: (1) the announcement of Freddie Gray's death while in coma at 7:00am on April 19, (2) the looting of a 7-Eleven convenience store around 8:00pm on April 25, (3) the looting and burning down of a CVS pharmacy around 5:00pm on April 27, (4) the announcement of the curfew at 8:00pm on April 27, (5) the begin of the curfew at 10:00pm on April 28, (6) and a the announcement that Gray's death was ruled a homicide by the involved police officers at 10:30 on May 1. 

We draw the following conclusions from these figures:
\begin{itemize}
\item As for the Baltimore events, there is a tight overlap between keywords' dynamic centrality measures and emergent events, with each dynamic measure sharply increasing in response to the event..
\item Dynamic measures are again superior to their static versions, with only one or two significant spikes that correspond to the ground truth events.
\item Spikes in simple word frequency overlap well with the emergent events. Movement in the time series, however, is very erratic, with spikes occurring throughout the entire time period, which diminishes the use of frequency counts for correctly identifying the onset of emergent events.
\end{itemize}

\begin{figure*}
\begin{center}
\includegraphics[width=\textwidth]{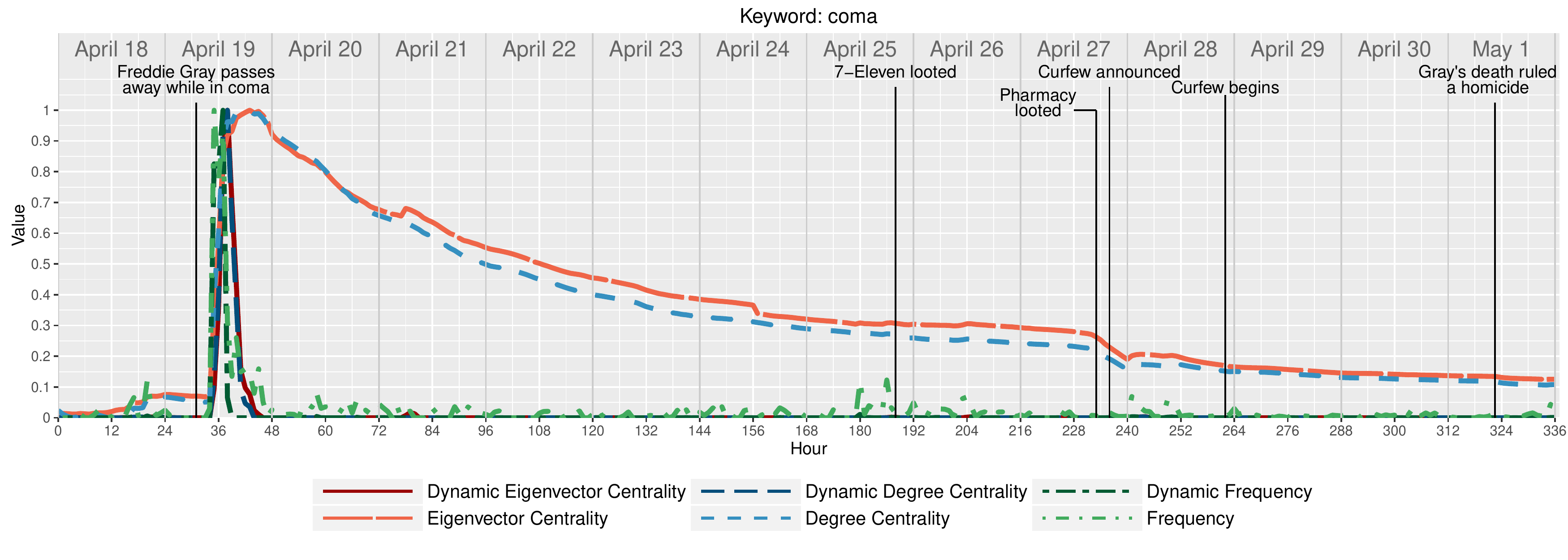}
\caption{Timeline of actual events during Baltimore protests together with importance measures for keyword ``coma''. \label{fig:Baltimore_timeline_coma}}
\end{center}
\end{figure*}

\section{Conclusions}

We presented a generalization of static semantic graph frequency, degree and eigenvector centrality measures into corresponding dynamic versions with applications in emergent keyword detection and data stream summarization. The proposed methods have been extensively evaluated on real-life streams of tweets associated with public safety events. We observed that our novel dynamic centrality indices successfully detect emergent keywords and provide concise and meaningful summarization. One of the promising future research directions is adapting these methods into smooth stream processing and summarization (instead of the discretized one) in which the summary elements will not be repeated in the next few time segments if a similar information has been detected and summarized in the previous time step. The implementation of the proposed approach is available at \cite{dyncentr-impl}.


\section*{Acknowledgments}
We thank the Clemson CBBS One-Year accelerate grant program for providing funds for this research.



\bibliographystyle{plain}
\bibliography{fullbib-ilya,ilya,alex,yuheng}
%



\end{document}